\documentclass[prb,twocolumn,noshowpacs]{revtex4}
\usepackage{amsmath,amssymb,graphicx,color}

\input{tcilatex}

\begin{document}

\title{Cavitation-induced ignition of cryogenic hydrogen-oxygen fluids}
\author{V. V. Osipov$^{1,2}$, C. B. Muratov$^{3}$, E. Ponizovskya-Devine$%
^{1,2}$, M.Foygel$^{4}$, V. N. Smelyanskiy$^{1}$}
\affiliation{$^{1}$Intelligent Systems Division, D\&SH Branch, NASA Ames Research Center,
MS 269-1, Moffett Field, CA 94035}
\affiliation{$^{2}$Mission Critical Technologies, Inc., 2041 Rosecrans Avenue, Suite 225,
El Segundo, CA 90245}
\affiliation{$^{3}$Department of Mathematical Sciences, New Jersey Institute of
Technology, Newark, NJ 07102 \\
$^{4}$South Dakota School of Mines and Technology, Rapid City, SD, 57701}
\date{\today }

\begin{abstract}
The Challenger disaster and purposeful experiments with liquid hydrogen (H2)
and oxygen (Ox) tanks demonstrated that cryogenic H2/Ox fluids always
self-ignite in the process of their mixing. Here we propose a
cavitation-induced self-ignition mechanism that may be realized under these
conditions. In one possible scenario, self-ignition is caused by the strong
shock waves generated by the collapse of pure Ox vapor bubble near the
surface of the Ox liquid that may initiate detonation of the gaseous H2/Ox
mixture adjacent to the gas-liquid interface. This effect is further
enhanced by H2/Ox combustion inside the collapsing bubble in the presence of
admixed H2 gas.
\end{abstract}

\maketitle

The source for the formation of flames in the cryogenic hydrogen/oxygen
(H2/Ox) fuel mixture during the Challenger disaster in 1986 still remains a
mystery. The fireball which caused the orbiter's destruction appeared near
the ruptured intertank section between the liquid H2 (LH2) and liquid Ox
(LOx) tanks, but not near the hot jets from the nozzles \cite{Chal-1,Chal-2}%
. Purposeful experiments with LOx and LH2 tanks carried out by NASA\cite%
{HOVI} showed that cryogenic H2/Ox mixtures always self-ignite when the
flows of cryogenic fluids containing gaseous hydrogen (GH2), gaseous oxygen
(GOx) and LH2 mix with a turbulent LOx stream. Since this effect can lead to
catastrophic events, understanding its mechanisms is a problem of great
importance.

In this Letter we propose a cavitation-induced self-ignition mechanism of
cryogenic H2/Ox fluids. Cavitation is the formation and compression of vapor
bubbles in flowing liquids driven by abrupt pressure variations. Due to
inertial motion of the liquid this process leads to a rapid collapse of the
bubbles and spiking of the gas temperature and pressure inside the bubbles,
producing strong shock waves\cite{Bren,Fu}. Here we discuss possible
scenarios of cavitation-induced ignition in cryogenic Ox/H2 fluids. We
concentrate on the most transparent scenario related to the collapse of a
vapor bubble in the Ox liquid near the interface between LOx and the GH2/GOx
mixture.

\begin{figure}[b]
\centering
\includegraphics[width=3.5in]{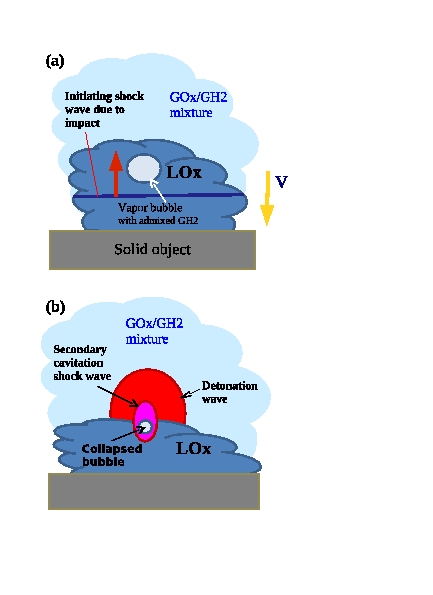}
\caption{A scenario of cavitation-induced ignition of GOx/GH2 mixture: (a)
an initiating \textquotedblleft weak\textquotedblright\ shock wave in a LOx
blob (with a bubble) forms due to its impact with a solid object; (b)
collapse of the bubble near the liquid-gas interface under the action of the
initiating shock wave and generation of a strong secondary shock wave
inducing detonation of the GH2/GOx mixture.}
\label{fig1}
\end{figure}

\begin{figure}[t]
\centering\includegraphics[width=3.5in]{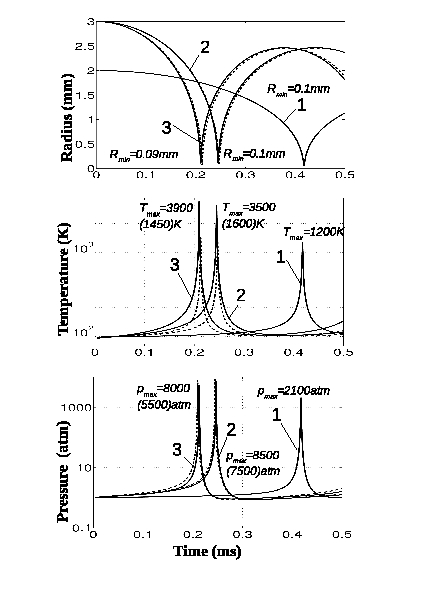}
\caption{Bubble collapse dynamics for the initial total gas pressure $%
p_{0}=1 $ atm. Other initial parameters are: radius $R_{0}=2$ mm,
overpressure of the initiating shock wave $\Delta p=0.25$ atm, GH2 partial
pressure $p_{H2}=0 $ (curve 1); $R_{0}=3$ mm, $\Delta p=1.5$ atm, $%
p_{H2}=0.01$ atm (curve 2); $R_{0}=3$ mm, $\Delta p=2.0$ atm, $p_{H2}=0.015$
atm (curve 3). The dashed lines and the values in parentheses are obtained
without considering combustion in the bubble. }
\label{fig2}
\end{figure}

\begin{figure}[t]
\centering\includegraphics[width=3.5in]{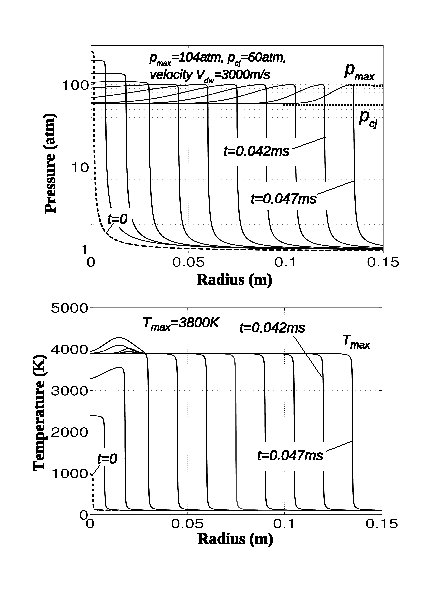}
\caption{Formation of cavitation-induced hemispheric detonation wave in
stoichiometric gaseous H2/Ox mixture with temperature $T=100$K and pressure $%
p=1$ atm. The initial conditions are: temperature $T=1000$K and pressure $%
p=250$ atm in the central area with radius $r\leq 0.15$ mm (dashed curves). $%
p_{cj}$ - steady-state pressure in the detonation wave (Chapman-Jouguet
pressure\protect\cite{cantera})}
\label{fig3}
\end{figure}

Vapor bubbles, most likely with admixed GH2, can be created in the falling
LOx blobs as a result of mixing of gaseous H2 and Ox with a turbulent stream
of liquid Ox. A pressure jump between LOx and the bubbles may arise, for
example, due to shock waves arising as a result of an impact of a LOx blob
against a solid object (Fig.\ref{fig1}a). The overpressure in such a shock
wave is of order $\Delta p\simeq \tfrac{1}{2}\rho _{L}v^{2}\gtrsim 2$ atm
even for moderate velocities $v\gtrsim 20$ m/s of the liquid. Such a
\textquotedblleft weak\textquotedblright\ shock wave cannot induce ignition
of the GH2/GOx mixture directly, but it can initiate cavitation collapse of
the vapor bubbles inside LOx. The below computations show that such weak
initiating shock waves can lead to the formation of bubbles of a small
radius $R_{\min }\simeq 0.1$ mm with huge pressures $p\gtrsim 1000$ atm and
temperatures $T\gtrsim 2500$K inside (Fig. \ref{fig2}). This causes ignition
of the GOx/GH2 mixture inside the bubble. A strong secondary shock wave
generated by the cavitation bubble collapse near the LOx interface may then
propagate into the gaseous H2/Ox mixture next to the LOx interface (Fig. \ref%
{fig1}b). We demonstrate that such a localized shock wave is sufficient to
induce detonation in cryogenic GH2/GOx mixtures (Fig. \ref{fig3}).

To simulate the bubble collapse, we used the standard fluid dynamics
equations describing conservation of the mass, energy and momentum under the
assumption of spherical symmetry\cite{Landau,Bren}. We treated the liquid as
incompressible and inviscid, neglected surface tension at the interface,
treated the gas phase as a mixture of ideal gases and took into account
diffusion and thermodiffusion of the admixed GH2 (see supplementary material
for more detail). We also modeled combustion inside the bubble, using a
simplified model based on the assumption that the burning rate is limited by
the initiation reactions $\mathrm{H}_{2}+\mathrm{O}_{2}\rightarrow \mathrm{O}%
\mathrm{H}+\mathrm{O}\mathrm{H}$ and $\mathrm{H}_{2}+\mathrm{O}%
_{2}\rightarrow \mathrm{H}\mathrm{O}_{2}+\mathrm{H}$, which have the lowest
rates. Thus, we modeled the GH2/GOx combustion by a brutto reaction $\mathrm{%
H}_{2}+\mathrm{O}_{2}\rightarrow \mathrm{H}_{2}\mathrm{O}+\tfrac{1}{2}%
\mathrm{O_{2}}$ with the rate $G_{comb}(T)=c_{H2}c_{Ox}[1.1\cdot 10^{8}\exp
(-19680/T)+1.48\cdot T^{2.433}\exp (-26926/T)]$ m$^{3}$/(mol$\cdot $sec),
where $c_{H2}$ and $c_{Ox}$ are the molar concentrations of GH2 and GOx,
respectively, and $T$ is in degrees Kelvin \cite{Data}. We note that this
approximation is close to the one-step mechanism of Mitani and Williams \cite%
{Wil}. Also, the model predicts the same steady detonation wave parameters
as those obtained\cite{HOVI} with the help of the model taking into account
17 main chain reactions of GOx/GH2 mixture combustion\cite{cantera}.

Our simulations show that under the action of initiating shock wave with
overpressure $\Delta p\geq 0.15$ atm the maximum pressure and temperature in
pure vapor bubbles of initial radius $R_{0} \gtrsim 2mm$ collapsing in LOx
exceed $1500$ atm and $800$K, respectively, when the bubble radius reaches
its minimum value of $R_\mathrm{min}\simeq 0.1$ mm (curve 1 in Fig.2). With
the increase of $\Delta p$ the values of $p_\mathrm{max}$ and $T_\mathrm{max}
$ increase and the value of $R_\mathrm{min}$ decreases ($R_\mathrm{min}\leq
0.05$ mm at $\Delta p\geq 0.5$ atm). The lower the value of $R_\mathrm{min}$%
, the harder it is to achieve ignition. The presence in the bubble of even a
small amount of non-condensable GH2 sharply decreases the values of $p_%
\mathrm{max}$ and $T_\mathrm{max}$. However, at large enough $\Delta p \geq
0.5$ atm and $R_{0}\gtrsim 2$ mm the bubble collapse leads to high
temperatures and ignition of the Ox/H2 mixture in the bubble. As a result of
this local explosion the values of $p_\mathrm{max}$ and $T_\mathrm{max}$ in
the bubble reach gigantic values $p_\mathrm{max}\geq8000$ atm and $T_\mathrm{%
max}\geq3500$K (Fig. 2). We note that in reality the super-hot and
super-compressed O, H, OH species forming in the process of GH2/GOx
combustion\cite{cantera} inside the bubble may be ejected into the space
above the LOx surface and easily ignite the GH2/GOx mixture nearby.

The same equations of gas dynamics used in the cavitation simulations were
also employed to analyze the ignition of GH2/GOx mixtures by a localized
strong shock wave generated by the collapsing bubble. We envision a
hemispherical shock wave propagating in the unconfined GH2/GOx mixture above
the LOx-gas interface (see Fig. \ref{fig1}b), initiated by a local increase
in gas pressure and temperature within the radius $R_{0}\sim R_{\mathrm{min}}
$. We see that the local jump of the pressure $p\gtrsim 200$ atm and
temperature $T\gtrsim 500$K in the GH2/GOx mixture area of radius $%
R_{0}\gtrsim 0.1$mm is sufficient to induce detonation in stoichiometric
GH2/GOx mixtures (Fig. 3).

We now list other possible cavitation-induced scenarios that may lead to
self-ignition of cryogenic H2/Ox fluids upon mixing:

\begin{enumerate}
\item[(i)] Formation of rarefied-vapor bubbles in LOx.

\item[(ii)] Formation of GOx bubbles in LOx with a thin chilled layer near
the bubble surface.

\item[(iii)] Injection of cold LH2 droplets into ``hot'' LOx.

\item[(iv)] Injection of ``hot'' LOx droplets into cold LH2.
\end{enumerate}

In scenario (i), rarefied-vapor bubbles form in the LOx stream impinging on
an obstacle and collapse under the liquid pressure. In scenario (ii) bubbles
may form as a result of an impact of two large LOx blobs whose surfaces were
chilled by contact with very cold ($T\simeq 20$K) surrounding GH2. Due to
the low near-surface temperature the pressure in the bubble may quickly drop
because of intense vapor condensation, leading to a rapid bubble collapse.
In scenario (iii) the pressure inside the bubble will quickly grow and may
become much greater than the pressure $p_{L}$ in the liquid bulk. As a
result, the bubble radius will increase and, due to the inertial motion of
the liquid, the pressure in the bubble can become much less than $p_{L}$. As
a consequence, the bubble will start to collapse, and the gas temperature
and pressure inside the bubble may achieve very high values, initiating a
local explosion and a strong shock wave. Finally, in scenario (iv) heavy
droplets of LOx may penetrate deeply into LH2 (a light fluid), causing
intense evaporation of LH2 and formation of a GH2/GOx bubble inside LH2 that
will grow in size and then collapse due to inertial motion of the liquid.
Since the critical temperature $T_{c}=33.2$K of H2 is significantly below
the freezing temperature $T_{m}\simeq 54$K of LOx, the evaporation of LH2 in
contact with LOx may acquire an explosive character, resulting in even more
dramatic outcomes.

To summarize, we have identified a possible mechanism of ignition in
cryogenic H2/Ox fluids which relies on the generation of strong shock waves
by the cavitational collapse of vapor bubbles close to the liquid-gas
interface in the process of cryogenic H2/Ox mixing. We showed that the
presence of LOx blobs surrounded by GH2/GOx mixture may be sufficient to
initiate H2/Ox ignition, including strong detonation waves. We further
proposed several other scenarios that include mixing of LH2 with LOx and
resulting in even more dramatic consequences. More detailed studies of these
mechanisms are currently underway. Finally, we note that the\ proposed
self-ignition mechanisms should be very important for understanding
conditions and risks of explosion in cryogenic H2/Ox-based liquid rockets
and other space vehicles.

The work of CBM was supported by NASA via grant NNX10AC65G.

\section*{APPENDIX: GOVERNING EQUATIONS}

Neglecting the surface tension and treating the liquid as incompressible and
inviscid, the equations for the liquid phase may be reduced (see e.g.\cite%
{Bren}) to a single ordinary differential equation for the bubble radius $R$%
: 
\begin{eqnarray}
&&R{\frac{d^{2}R}{dt^{2}}}+\frac{3}{2}\left( {\frac{dR}{dt}}\right) ^{2}+%
\frac{j}{\rho _{L}}{\frac{dR}{dt}}+\frac{R}{\rho _{L}}{\frac{dj}{dt}}  \notag
\\
&=&\frac{p_{m}-p_{L0}}{\rho _{L}}+\frac{(2\rho _{L}-\rho _{v})j^{2}}{2\rho
_{v}\rho _{L}^{2}},
\end{eqnarray}%
and the advection-diffusion equation for the liquid temperature $T_{L}$: 
\begin{eqnarray}
&&\frac{\partial T_{L}}{\partial t}+\left( \frac{R}{r}\right) ^{2}\left( {%
\frac{dR}{dt}}+\frac{j}{\rho _{L}}\right) \frac{\partial T_{L}}{\partial r} 
\notag  \label{eq:2} \\
&=&{\frac{\kappa _{L}}{c_{L}\rho _{L}}}\ {\frac{1}{r^{2}}}{\frac{\partial }{%
\partial r}}\left( r^{2}{\frac{\partial T_{L}}{\partial r}}\right) .
\end{eqnarray}%
Here $r\geq R(t)$ is the radial coordinate, $\rho _{L}$, $c_{L}$ and $\kappa
_{L}$ are the liquid density, specific heat and thermal conductivity,
respectively, $p_{m}$ and $\rho _{v}$ are the pressure and the vapor mass
density, respectively, in the gas mixture at the liquid-gas interface, $%
p_{L0}$ is the liquid pressure far from the bubble. The vapor condensation
flux $j$ is given by the well-known Hertz-Knudsen equation\cite{Bren}: 
\begin{equation}
j=\frac{\alpha (p_{v}-p_{s}(T_{s}))}{\sqrt{2\pi R_{v}T_{s}}},\qquad
p_{s}(T)=p_{c}\left( T_{s}/T_{c}\right) ^{\lambda },  \label{cond}
\end{equation}%
where $T_{s}$ and $p_{v}$ are the vapor temperature and pressure,
respectively, at the liquid-gas interface, $R_{v}$ is the vapor gas
constant, $\alpha $ is the accommodation coefficient. Here we used a simple
approximation for the dependence of the saturated vapor pressure $%
p_{s}(T_{s})$ on the liquid-gas interface temperature $T_{s}$, where $p_{c}$
and $T_{c}$ are critical pressure and temperature, respectively, of the
vapor, and $\lambda $ is a dimensionless parameter (see \cite{OxLamda}).
Note that $p_{s}$ depends strongly on $T_{s}$, which significantly affects
the bubble dynamics \cite{Fu,Os2000}.

In the gas phase, we have the conservation of momentum and energy (again
neglecting viscosity effects): 
\begin{eqnarray}
\frac{\partial u_{m}}{\partial t}+u_{m}\frac{\partial u_{m}}{\partial r} = - 
\frac{1}{\rho _{m}}\frac{\partial p_{m}}{\partial r}, \\
{\frac{\partial E_m}{\partial t}+{\frac{1 }{r^2}} \frac{\partial }{\partial
r } \Bigl( r^2 u_m ( p_{m}+E_m) \Bigr) }  \notag \\
= {\frac{1 }{r^2}} {\frac{\partial }{\partial r}\left( \varkappa _{m} r^2 
\frac{\partial T_{m}}{\partial r}\right) +Q_{h}}G_{comb}.
\end{eqnarray}
Here $\rho_m$, $u_m$, $p_m$, $T_m$, $E_m$ are the density, radial velocity,
pressure, temperature, and energy density, respectively, of the gas mixture
located at $r \leq R(t)$, and $\varkappa _{m}=\left( \frac{T_m}{T_{0}}%
\right) ^{1/2}\sum_i c_{i} \varkappa _{i}^0 / \left( \sum_i c_{i} \right)$
is the thermal conductivity of the gas mixture as a function of $T_m$, the
molar concentrations $c_i = \{ c_{H2}, c_{Ox}, c_{H2O} \}$ of different
molecular components and thermal conductivities of pure gas species $%
\varkappa_i^0$ at some reference temperature $T_0$. Finally, $G_{comb}$ is
the combustion rate and $Q_h$ is the combustion heat. Treating all the gas
species as diatomic ideal gases for simplicity, we have 
\begin{eqnarray}  \label{eq:1}
p_{m}=R_g T_{m}c _m, \qquad {E}{=\frac{5}{2} R_g T_{m}c_{m}+\frac{1}{2}\rho
_{m}u_{m}^{2}, }
\end{eqnarray}
where $c_{{m}} =\sum_{i}c_{i}$ and $\rho_{m}=\sum_{i}c_{i}M_{i}$ are the
total molar concentration and mass density, respectively, of the gas
mixture, with $R_g$ the universal gas constant and $M_i$ the molar masses of
the gas species. Note that we kept the kinetic energy term in the expression
for $E_m$ in order to be able to account for possible rapid onset of
combustion inside the bubble.

The dynamics of the GH2/GO2 mixture combustion are described by the
continuity equations for the molar concentration of the mixture components $%
c_{i}$: 
\begin{eqnarray}
\frac{\partial c_{H2O}}{\partial t}+\frac{1}{r^{2}}\frac{\partial }{\partial
r}(r^{2}c_{H2O} u_{m})=G_{comb}, \\
\frac{\partial c_{Ox}}{\partial t}+\frac{1}{r^{2}}\frac{\partial }{\partial
r }(r^{2}c_{Ox}u_{m})=-\tfrac12 G_{comb}, \\
{\frac{\partial c_{H2}}{\partial t}+\frac{1}{r^{2}}\frac{\partial }{\partial
r}(r^{2}c_{H2}u_{m})} + G_{comb}  \notag \\
{=}{\frac{1}{r^{2}}\frac{\partial }{\partial r} \left\{ r^{2}D_{H2} \left( 
\frac{\partial c_{H2}}{\partial r}-\frac{c_{H2}}{2T_{m}} \frac{\partial
T_{m} }{\partial r} \right) \right\},}
\end{eqnarray}
where ${D_{H2}=\left( T_{m} / T_{0} \right) ^{3/2}\left( p_{0} / p_{m}
\right) D_{H2}(T_{0},p_{0})}$ (see \cite{fizkiny}). Note that in the last
equation we included the effect of GH2 diffusion, which may be significant
due to high diffusivity of hydrogen and its role as a non-condensable gas
during the bubble collapse.

Finally, the boundary and the initial conditions for the equations above
are: 
\begin{eqnarray}
\left. \frac{\partial T_{m}}{\partial r}\right\vert _{r=0} &=&0,\ \left.
\left( \kappa _{L}\frac{\partial T_{L}}{\partial r}-\kappa _{m}\frac{%
\partial T_{m}}{\partial r}\right) \right\vert _{r=R}=jq_{h},  \notag \\
\left. T_{m}\right\vert _{t=0} &=&\left. T_{L}\right\vert
_{t=0}=T_{L0},\quad \left. T_{L}\right\vert _{r=R}=\left. T_{m}\right\vert
_{r=R}=T_{s},  \notag \\
\left. T_{L}\right\vert _{r\gg R} &=&T_{L0},\quad \left. \frac{\partial c_{i}%
}{\partial r}\right\vert _{r=0}=0,\quad \left. c_{i}\right\vert
_{t=0}=c_{i}^{0},  \notag \\
\left. u_{m}\right\vert _{r=0} &=&0,\quad \left. u_{m}\right\vert _{r=R}={%
\frac{dR}{dt}}-{\frac{j}{M_{Ox}c_{Ox}}},\quad \left. u_{m}\right\vert
_{t=0}=0,  \notag \\
\left. {\frac{jc_{H2}}{M_{Ox}c_{Ox}}}\right\vert _{r=R} &=&\left.
D_{H2}\left( {\frac{\partial c_{H2}}{\partial r}}-{\frac{c_{H2}}{2T_{m}}}{%
\frac{\partial T_{m}}{\partial r}}\right) \right\vert _{r=R},  \notag \\
R(0) &=&R_{0},\quad {\frac{dR(0)}{dt}}=0,\qquad
\end{eqnarray}%
where $q_{h}$ is the latent heat of LOx vaporization. Additional conditions
are presented in (Fig. \ref{fig2}).

The simulations were done using Godunov's scheme with variable time-step
for stiff problem. The time step was varied depending on the maximum of the
time derivatives of bubble radius, temperature and pressure. In particular
we use Monotone Upstream-centered Schemes for Conservation Laws (MUSCL)
based numerical scheme that extends the Godunov's scheme idea of linear
piecewise approximation to each cell by using slope limited left and right
extrapolated states.

List of all constants used in the simulations are represented in Table I.

\begin{table}[tbp]
\caption{Parameters of oxygen and hydrogen in SI units used in the
simulations}
\centering
\begin{tabular}{|c|c|c|l|}
\hline\hline
Parameter & Oxygen & Hydrogen & Meaning  \\[0.5ex] \hline
$c_L$ & $1700$ & - & specific heat of the liquid   \\ 
$\rho_L$ & $1141$ & - & liquid density   \\ 
$V_L$ & $1.88\cdot10^{-4}$ & - &kinematic viscousity of\\
& & & liquid   \\ 
$C_0$ & $1130$ & - & sound velocity (T=90K)  \\ 
$T_c$ & $154.58$ & - & critical temperature  \\ 
$p_c$ & $5.043\cdot10^6$ & - & critical pressure  \\ 
$\lambda$ & $7$ & - & factor   \\ 
$q_h$ & $2.13\cdot10^5$ & - & specific heat of evaporation \\ 
$C_v$ & $653$ & $10130$ & specific heat of vapor \\
 & & & at V=const,T=300K  \\ 
$R_v$ & $264$ & $4124$ & gas constant  \\ 
$C_p$ & $917$ & $14270$ & specific heat of vapor\\
 & & & at p=const,T=300K  \\ 
$\gamma$ & $1.4$ & $1.4$ &   \\ 
$\kappa_{i}$ & $0.0565$ & $0.0165$& 
thermal conductivity of\\
 & & &saturated vapor (T=80K) \\ 
$\kappa_{L}$ & $0.17$ & - & thermal conductivity\\
 & & & of liquid (T=80K)   \\ 
$Q_{h}$ & $1.418\cdot10^8$ & - & heat of combustion  \\ 
$\sigma$ & $7.3\cdot10^{-3}$ & - & liquid surface tension  \\ 
$D_{i}$ & - & $6.5\cdot10^{-5}$ & diffusion constant\\
 & & & (T=270K, p=1atm) \\[1ex] \hline
\end{tabular}
\label{table:oxygen}
\end{table}


\end{document}